\documentclass[aps,twocolumn]{revtex4}

\usepackage{epsf}

\newcommand{\be}{\begin{eqnarray}}
\newcommand{\ee}{\end{eqnarray}}

\newcommand{\GeV}{\hbox{GeV}}
\newcommand{\fm}{\hbox{fm}}
\newcommand{\gs}{g_{\rm s}}
\newcommand{\mD}{m_{\rm D}}
\newcommand{\q}{{\bf q}}
\newcommand{\p}{{\bf p}}
\newcommand{\F}{{\bf F}}
\newcommand{\h}{{\bf h}}
\newcommand{\eqn}[1]{Eq.~(\ref{#1})}

\begin{document}

\preprint{McGill-Nuc}

\title{Energy Loss of Leading Partons
in a Thermal QCD Medium}

\author{Sangyong Jeon$^{\dagger, *}$}
\author{Guy D.\ Moore$^{\dagger}$}

\affiliation{$^\dagger$Physics Department, McGill University, 3600
University Street, Montreal, QC H3A-2T8, Canada}%
\affiliation{$^*$RIKEN-BNL Research Center, Brookhaven National
Laboratory, Upton, NY 11973}

\begin{abstract}
We consider bremsstrahlung energy loss for hard partons traversing a
quark-gluon plasma.  Accounting correctly for the probabilistic nature of
the energy loss, and making a leading-order accurate treatment of
bremsstrahlung, we find that the suppression of the spectrum is nearly
flat, with the most suppression at energies $E \sim 30 T$ ($T$ the QGP
temperature), in contrast to previous literature but in agreement with
experimental data.  This flat pattern should also be observed at the LHC.

\end{abstract} 

\maketitle

\noindent {\bf Introduction}

 In highly relativistic heavy ion collisions, production of hard $p_T$
 partons precedes most other processes, simply because the time
 scale of the production is short, $\tau \sim 1/p_T$.  In particular, the
 production of hard secondary partons precedes the formation of a
 quark-gluon plasma (QGP).  
 Therefore, the produced hard partons find themselves in an environment far
 different than the vacuum. The interaction between the parton
 and the environment influences the final spectrum of high $p_T$
 hadrons in a non-trivial way.  In particular, if the QGP is very dense,
 we expect energy loss, leading to an energy dependent suppression of
 the high $p_T$ spectrum, a phenomenon called ``jet quenching.''  The
 extent of jet quenching can be used to learn about the QGP
 \cite{Gyulassy:2003mc}.
 Experimentally, the CERN SPS gave little evidence of jet
 quenching, but RHIC has seen a rather dramatic suppression of high
 $p_T$ pions.
 
 The results from RHIC 
 experiments\cite{Adams:2003im,Adler:2003qi,Back:2003qr,Arsene:2003yk}
 are surprising in many ways.  
 At $p_T \sim 3\,\GeV$, we already see a substantial
 suppression.  Furthermore, the suppression continues more or less in constant
 fashion up to the highest $p_T$ measured so far $(\sim 10\,\GeV)$.
 The amount of suppression is also rather big; the ratio of high $p_T$ events
 to the number expected based on proton-proton data is $R \simeq 1/5$. 
 
 Theoretically, it is well established that the main energy loss
 mechanism of a fast parton is the bremsstrahlung of gluons in medium.
 The strength of the bremsstrahlung in medium depends on a coherence 
 effect called the Landau-Pomeranchuk-Migdal
 (LPM) effect.  The first quantitative treatment of this effect (in QED)
 was by Migdal \cite{Migdal}; more recently it
 has been considered in QCD by several authors
 \cite{Gyulassy:1993hr,BDMPS,BDMS,Zakharov}. 

\begin{figure}[ht]
 \vspace{0.1in}
  \begin{center}
  \epsfxsize=0.45\textwidth
  \epsfbox{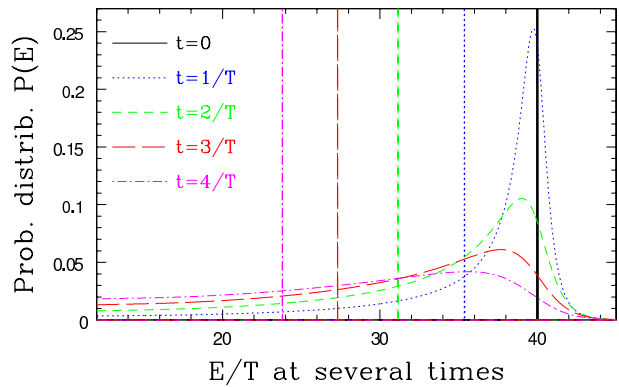}
  \end{center}
  \vspace{-0.15in}
  \caption{Time evolution of an initially monoenergetic ensemble of
  quarks.  For comparison, vertical bars show the energy of the quarks
  if we take the energy loss to be a steady process, determined by 
  \eqn{eq:dEdt}.
  }
  \label{fig:evolve_40}
 \end{figure}

 Jet quenching through bremsstrahlung energy loss has been considered by
 several authors before
 \cite{Gyulassy:1993hr,BDMPS,BDMS,Gyulassy:2000er,BDMS2001,
 Wang:2001cy,Sarcevic}.  
 The purpose of this paper is to revisit the energy loss calculation,
 particularly emphasizing two points.  
 In many previous treatments, the (path length
 dependent) average energy loss is computed and applied to all primary
 partons.  However, bremsstrahlung is not well described by an (path
 length dependent) energy loss. Bremsstrahlung energy loss is dominated
 by hard emissions; for any time scale up to the shower length, a parton
 typically has made about half its energy loss in a single emission.
 Therefore, if a sample of partons initially has
 the same energy, then after going through some pathlength of medium,
 the distribution of final energies is as broad as the mean energy
 loss.  This is illustrated in Fig.~\ref{fig:evolve_40}.
 This is especially important when the initial energy distribution is a
 rapidly falling function of energy; the final distribution is dominated by the
 few partons which happen not to lose much energy.
 This has previously been emphasized by Baier {\it et.\
 al.}\ \cite{BDMS2001}, who found that it significantly influences the
 final spectrum (see also \cite{Gyulassy:2001nm}).  
 To account for it, we directly evolve the
 spectrum of the partons as they undergo bremsstrahlung energy loss.

 Second, most previous treatments have taken the LPM effect to be a
 parametrically large suppression.  This is formally true whenever the
 parton and the emitted gluon are very energetic, $E_{\rm parton},
 E_{\rm gluon} \gg T$.  
 However, the LPM suppression (actual rate over Bethe-Heitler rate) is
 only a factor of $1/2$ at $E_{\rm gluon}=10 T$;
 for less energetic gluons the LPM suppression is
 small (and approximations which take it to be large are badly in error).  
 When the parton spectrum is steeply falling, the most important
 bremsstrahlung events have $E_{\rm gluon} \ll E_{\rm parton} $;
 so at realistic parton energies one cannot take the LPM effect to be
 {\em parametrically} large.  Therefore, we will use
 the formalism developed by 
 Arnold, Moore and Yaffe (AMY) \cite{AMY245},
 which correctly treats the LPM effect (up to $O(g_{\rm s})$
 corrections) at all  energies $E_{\rm gluon} > g_{\rm s} T$.

In this short paper we will concentrate on the qualitative features of the
spectrum, specifically its shape (the $p_T$ dependence of $R$).  The goal 
is to show that the trend seen in the data--that $R$ falls 
slightly and levels off, but does not rise with $p_T$ at least at accessible
momenta--is the trend expected from bremsstrahlung energy loss.  In previous
literature it has appeared necessary to explain the data by invoking additional
many body effects such as the Cronin effect and shadowing
 \cite{Vitev:2002pf,BDMS2001,Muller:2002fa}.
 In particular, we predict that
 even at LHC, the flat suppression pattern seen in
 RHIC experiments should persist.
Since we are only after these qualitative features, we will simplify the
treatment somewhat, and consider a  static thermal medium of quarks and gluons
 at a  constant temperature $T > T_c$
 To be more quantitative, we need to take
 into account the nuclear geometry and the expanding system.  We will do
 so in a subsequent publication.  

\bigskip
\noindent {\bf Brief Description of Formalism}

We consider a small number of high energy partons traversing a
thermalized QGP.  The high energy partons are rare enough
that their dominant interactions are with the thermal bath particles.
We also work at leading nontrivial order in $\alpha_{\rm s}$.  This is
obviously an idealization, but it is hard to see how to do better at
present. 

A parton traversing the QGP undergoes a series of soft scatterings with
other constituents of the medium, with leading-order cross-section
\begin{equation}
\sigma_{\rm soft} = C_{\rm s} \gs^2 \int \frac{d^2 \q_\perp}{(2\pi)^2}
        C(\q_\perp) \, .
\end{equation}
(The group Casimir $C_{\rm s}$ is $C_f{=}\frac 43$ [quarks] or
$C_{\rm A}{=}3$ [gluons].)
Here $C(\q_\perp)$ is the differential rate to exchange transverse (to
the parton) momentum $\q_\perp$.  In a hot thermal medium, its value at
leading order in $\alpha_{\rm s}$ is \cite{AGZ}
\begin{equation}
\label{eq:Cq}
C(\q_\perp) = \frac{\mD^2}{\q_\perp^2(\q_\perp^2{+}\mD^2)} \, ,
\quad
\mD^2 = \frac{\gs^2 T^2}{6} (2 N_{\rm c} {+} N_{\rm f}) \, .
\end{equation}

These frequent soft scatterings can induce collinear splitting
(bremsstrahlung) of the parton.  The time scale over which the parton
and bremmed gluon overlap, in the absence of other scatterings, is
$
\tau \sim \frac{x p}{\p_\perp^2} \sim \frac{x p}{g^2 T^2} \, ,
$
with $x$ the momentum fraction of the gluon and $p_\perp$ the momentum
component of the gluon perpendicular to the original parton.  When
$\sigma_{\rm soft} \tau$ is large, additional collisions typically occur
while the parton and gluon are still coherent; this can frustrate the
original emission.

This problem has been treated in the QCD context by BDMPS \cite{BDMPS}
and by Zakharov \cite{Zakharov}.  AMY have re-analyzed it with almost the
same conclusions \cite{AMY245}; we outline the physics and 
summarize their results.  The probability of
emission of a gluon of momentum $k$ is schematically
\begin{equation}
\int dk_\perp \left| \langle p-k;k| \, \int_t J_\mu^a G^{\mu a}_{\rm hard}(t) 
        \, | p \rangle \right|^2 \, ,
\end{equation}
with the effects of soft collisions implicitly included in the time
evolution.  The time for the $J \cdot G$ insertion in the amplitude and
its conjugate differ; to get the emission rate we must
integrate over the difference of these times,
\begin{equation}
\frac{d\Gamma}{dk dt}\! \sim \!\! \int \!\! dt' \langle p | 
J_\mu^a G^{\mu}_{a}(t') | p{-}k;k \rangle \langle p{-}k;k |
J_\nu^b G^{\nu}_b(0) | p \rangle.
\end{equation}
The problem is then to evolve $| p{-}k;k \rangle \langle p |$
between time $0$ and time $t'$.  Its evolution equation is similar to a
Boltzmann equation, but with an extra phase accumulating term because
the states $|p\rangle$ and $|p{-}k;k \rangle$ have different
energies.  Performing the time integration (taking the medium to be
uniform on the scale of the formation time), the complete expression for
the bremsstrahlung rate turns out to be,
\begin{eqnarray}
\label{eq:dGamma}
\frac{d\Gamma(p,k)}{dk dt} & = & \frac{C_s \gs^2}{16\pi p^7} 
        \frac{1}{1 \pm e^{-k/T}} \frac{1}{1 \pm e^{-(p-k)/T}} \times
        \nonumber \\ && \times
\left\{ \begin{array}{cc} 
        \frac{1+(1{-}x)^2}{x^3(1{-}x)^2} & q \rightarrow qg \\
        N_{\rm f} \frac{x^2+(1{-}x)^2}{x^2(1{-}x)^2} & g \rightarrow qq \\
        \frac{1+x^4+(1{-}x)^4}{x^3(1{-}x)^3} & g \rightarrow gg \\
        \end{array} \right\} \times \nonumber \\ && \times
\int \frac{d^2 \h}{(2\pi)^2} 2 \h \cdot {\rm Re}\: \F(\h,p,k) \, ,
\end{eqnarray}
where $x\equiv k/p$ is the momentum fraction in the gluon (or either
quark, for the case $g \rightarrow qq$).  
The factors $1/(1\pm e^{-k/T})$ are Bose stimulation or Pauli blocking
factors for the final states, with $-$ for bosons and $+$ for fermions. 
$\h \equiv \p \times
{\bf k}$ is the invariant describing the non-collinearity of the final
states; $\h$ lives in a two dimensional transverse space.
$\F(\h,p,k)$ is the solution of an integral equation describing how
$|p-k;k \rangle \langle p|$ evolves with time, due to collisions and the
energy difference of the two states;
\begin{eqnarray}
2\h = &&\hspace{-0.14in} 
        i \delta E(\h,p,k) \F(\h) + g^2 \int \frac{d^2 \q_\perp}{(2\pi)^2}
C(\q_\perp) \times \nonumber \\ && 
\times \Big\{ (C_s-C_{\rm A}/2)[\F(\h)-\F(\h{-}k\,\q_\perp)] 
        \nonumber \\ && \hspace{0.2in}
        + (C_{\rm A}/2)[\F(\h)-\F(\h{+}p\,\q_\perp)] 
        \nonumber \\ && \hspace{0.2in}
        +(C_{\rm A}/2)[\F(\h)-\F(\h{-}(p{-}k)\,\q_\perp)] \Big\} , 
\label{eq:integral_eq1}
        \\
\!\!\!\!\delta E(\h,p,k) &&\hspace{-0.14in} 
        = \frac{\h^2}{2pk(p{-}k)} + \frac{m_k^2}{2k} +
        \frac{m_{p{-}k}^2}{2(p{-}k)} - \frac{m_p^2}{2p} \, .
\label{eq:integral_eq2}
\end{eqnarray}
Here $m^2$ are the medium induced thermal masses, equal to $m_D^2/2$ for
a gluon and $C_f \gs^2 T^2/4 = \gs^2 T^2/3$ for a quark.  For the case
of $g\rightarrow qq$, the $(C_s-C_{\rm A}/2)$ term is the one with
$\F(\h-p\,\q_\perp)$ rather than $\F(\h-k \,\q_\perp)$.  

The treatment of BDMPS is the same, except that it 
uses $(q^2{+}\mD^2)^2$ in the denominator of
\eqn{eq:Cq}, and drops the mass terms in \eqn{eq:integral_eq2}.  These
errors are not numerically significant.  They also typically solve
\eqn{eq:integral_eq1} in a large $h$ approximation, valid for large
$p/T,k/T$ but unreliable for $k \leq 10 T$.

Next, we use these expressions to evolve the hard gluon distribution
$P_{\! g}(p,t=0)$ and the hard quark plus antiquark distribution 
$P_{\! q}(p,t=0)$ with time, as they traverse the medium.
The joint equations for $P_{\! q}$ and $P_{\! g}$ are
\begin{eqnarray}
\frac{dP_{\! q}(p)}{dt} & \!\!\!= \!\!\!\! & \!\int_k\!\! 
        P_{\! q}(p{+}k) \frac{d\Gamma^q_{\!gg}(p{+}k,k)}{dkdt}
        -P_{\! q}(p)\frac{d\Gamma^q_{\!gg}(p,k)}{dkdt} 
        \nonumber \\ && \quad
        +2 P_{\! g}(p{+}k)\frac{d\Gamma^g_{\!qq}(p{+}k,k)}{dkdt}
        \, , \nonumber \\
\frac{dP_{\! g}(p)}{dt} & \!\!\!= \!\!\!\! & \!\int_k \!\! 
        P_{\! q}(p{+}k) \frac{d\Gamma^q_{\!qg}(p{+}k,p)}{dkdt}
        {+}P_{\! g}(p{+}k)\frac{d\Gamma^g_{\!gg}(p{+}k,k)}{dkdt}
        \nonumber \\ && \;
        -P_{\! g}(p) \left(\frac{d\Gamma^g_{\!qq}(p,k)}{dkdt}
        + \frac{d\Gamma^g_{\!gg}(p,k)}{dkdt} \Theta(2k{-}p) \!\!\right) ,
\label{eq:Fokker}
\end{eqnarray}
where the $k$ integrals run from $-\infty$ to
$\infty$.  The integration range with $k<0$ represents absorption of
thermal gluons from the QGP; the range with $k>p$ represents
annihilation against an antiquark from the QGP, of energy $(k{-}p)$.
In writing \eqn{eq:Fokker}, we used
$d\Gamma^g_{\!gg}(p,k)=d\Gamma^g_{\!gg}(p,p{-}k)$ and similarly for
$g\rightarrow qq$; the $\Theta$ function in the loss term for $g
\rightarrow gg$ prevents double counting of final states.  Since
bremsstrahlung energy loss involves only small $O(\gs T/p)$ changes to the
directions of particles, these equations can be used for the momentum
distributions in any particular direction.  For a single initial hard 
particle, they can be viewed as Fokker-Planck equations for
the evolution of the probability distribution of the particle energy and
of accompanying gluons.

\bigskip
\noindent {\bf Numerical Evaluations}

We evaluate \eqn{eq:dGamma} by the impact parameter space method of
\cite{AGMZ}, on a grid of points $p,k$; \eqn{eq:Fokker} is then solved on
this momentum space grid by a second order Runge-Kutta algorithm.  The
errors are quadratic in the momentum discretization provided proper care
is taken with the small $k$ behavior of $d\Gamma/dkdt$.  Discretization
errors are under control and numerical costs are modest.

%

 To have a realistic $p_T$ spectrum for both peripheral and central
 collisions, we need a realistic initial $p_T$ spectrum.
 Here we will use a parameterization taken from
 Ref. \cite{Wang:2001cy},
 \be
 \left. {d N\over d^2 p_T} \right|_{\rm init} \approx 
 {C\over \left(p_0^2 + p_T^2\right)^5} \, ,
 \ee
 as the initial $p_T$ spectrum of hard partons.  
 Ref.\ \cite{Wang:2001cy} fitted this to
 $\sqrt{s} = 200\,\GeV$ $p\bar p$ data, and obtained $p_0 \sim 1.75$ GeV
 (consistent with the initial spectrum in Ref.\cite{BDMS2001}).

 \bigskip
 \noindent
 {\bf Results and Discussion}
 
 Consider first the evolution of a sample of quarks, all of momentum
 $p$ at $t=0$.  Since bremsstrahlung is a statistical process,
 the range of momenta broadens and the mean value falls.
 We show such an evolution in Fig.~\ref{fig:evolve_40}.
 The unit time step in the figure was $16/g^4 T$, which is $\simeq 1/T$ for
 $\alpha_{\rm s}=1/3$.
 The figure shows clearly that the momentum distribution broadens as
 fast as the mean value falls.
 For comparison, we also plotted the particle energy we would obtain if
 we took the energy loss process to be a steady $dE/dt$ energy loss,
 determined by
\begin{equation}
\mbox{Energy loss approx.:}\quad
\frac{dE}{dt} = \int dk \frac{d\Gamma^q_{\!qg}(p,k)}{dkdt} \, ,
\label{eq:dEdt}
\end{equation}
 in which case all the particles would have the same energy at any given
 time.  We see that this is a bad description of the real energy
 distribution. 

\begin{figure}[ht]
 \vspace{0.1in}
  \begin{center}
  \epsfxsize=0.45\textwidth
  \epsfbox{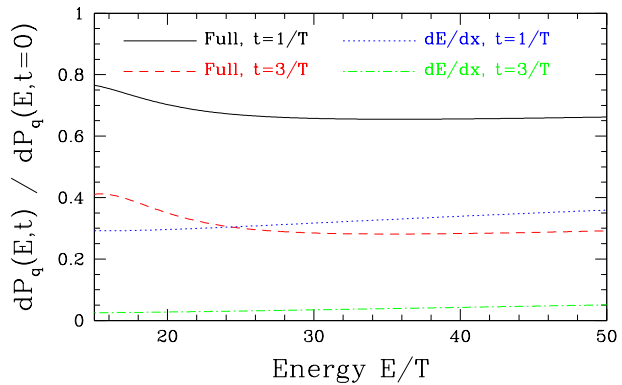}
  \end{center}
  \vspace{-0.15in}
  \caption{The ratios of the final and the initial momentum spectra for quarks.
  The black and red (solid) curves are calculated by solving 
  Eq.(\protect\ref{eq:Fokker}) directly.  The blue and green (dashed) 
  curves are calculated by first calculating $dE/dt$.}
  \label{fig:FullVsdEdt}
 \end{figure}

 In Fig.\ref{fig:FullVsdEdt}, we show the ratio of the final and the initial
 quark spectrum calculated in two ways.  
 The solid and dashed curves are the result
 of directly solving Eq.(\ref{eq:Fokker}), and the dotted and dash-dot
 curves are obtained by evolving particle energies according to
 \eqn{eq:dEdt}.  As the quarks move with
 almost the speed of light, we need not distinguish the time and the length.
 With the plasma temperature of about 0.4~GeV, the energy range shown
 corresponds to about 6~GeV to 20~GeV, with a time range of
 $0.5\,\fm$ to $1.5\,\fm$.

 {}From Fig.\ref{fig:FullVsdEdt}, it is obvious that the $dE/dt$ method
 is not a good approximation.  When
 particles with a steeply falling energy spectrum lose energy, the final
 spectrum is typically dominated by those few particles which happened
 not to lose much energy.  To give an extreme example, consider a
 spectrum falling as
 $p^{-10}$, and compare the effect of two energy loss mechanisms.  In
 the first, half the particles lose all their energy and half are
 unaffected; in the second, all particles lose exactly half their
 energy.  In the former case, the spectrum is reduced by a factor of
 $1/2$; in the latter case, it is reduced by $(1/2)^{10} \sim 1/1000$.
 Yet the mean energy loss of the two hypothetical
 processes is the same.  The real case is somewhere in between.  Almost
 all particles lose energy, but a few
 particles happen to suffer fairly small energy loss, while others suffer
 large energy loss.  This causes less suppression of the spectrum than if
 the energy loss process were uniform.  

\begin{figure}[t]
 \vspace{0.1in}
  \begin{center}
  \epsfxsize=0.45\textwidth
  \epsfbox{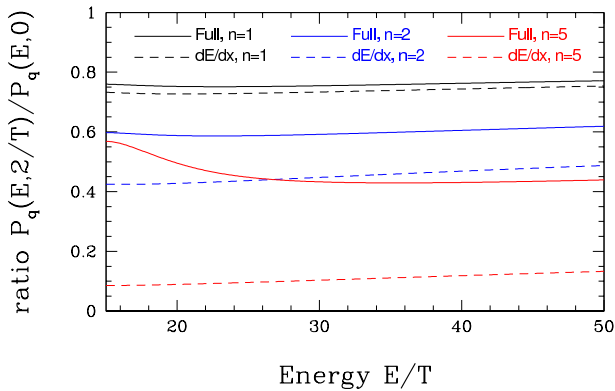}
  \end{center}
  \vspace{-0.15in}
  \caption{The ratios of the final and the initial momentum spectra for quarks.
  The solid curves are calculated by solving 
  Eq.(\protect\ref{eq:Fokker}) directly.  The dashed curves are
  calculated by first calculating $dE/dt$.  The evolution time is $2/T$.
  The integer $n$ corresponds to having initial spectrum 
  $f_0 \propto 1/(p_0^2 + p^2)^n$.}
  \label{fig:var_ratio}
 \end{figure}

 This effect becomes more important as the spectrum of initial particle
 energies becomes steeper.  We illustrate this in
 Fig.\ref{fig:var_ratio}, which shows the ratio $P_{\! q}(E,t=2/T)/P_{\!
 q}(E,t=0)$ for initial distributions of the form $C/(p_0^2 + p^2)^n$.
 For small values of $n$, the $dE/dt$ method is not that bad.
 However, as $n$ grows, the result of the 
 $dE/dt$ calculation deviates more and more from the correct one.

 Also note that the energy dependence of the suppression factor is different
 from the $dE/dt$ prediction; at $p/T\sim 15$ the suppression is
 actually becoming larger as $p$ increases, though eventually this
 behavior turns over.  This $p$ dependence is similar to what is observed
 experimentally in the ratio plot of $AA$ and $pp$ high $p_T$ spectra.
 As noticed before\cite{D'Enterria:2003rr,Muller:2002fa}
 the previous LPM effect calculations seem to be incompatible with 
 the current RHIC data essentially because these calculations show rising
 ratios while the data shows a flat or slightly
 decreasing ratio.  What we show here is that this discrepancy does not
 mean that the basic energy loss mechanism (LPM suppressed
 bremsstrahlung) is wrong.  Instead, the failure of most previous
 approximations is largely due to making the 
 $dE/dt$ approximation.  Part of the explanation also lies in the
 treatment of bremsstrahlung made here, in which the LPM effect is not
 taken as parametrically large, and absorption as well as radiation is
 allowed. 

 \bigskip
 \noindent
 {\bf Conclusion}
 
 In this work, we have demonstrated that LPM suppressed bremsstrahlung
 can in fact explain the qualitative features of the high $p_T$ 
 experimental data.  This is in contrast to previous findings. 
 This is because of two features of our treatment.  First, we
 determine the distribution of final energies an energetic parton can
 end up with, rather than using the mean.  This is important
 when the initial spectrum is steeply falling.  Second,
 we do not assume that the LPM effect is {\em parametrically} large from
 the beginning, but use a treatment which handles the transition between
 Bethe-Heitler and LPM correctly.  Since this transition occurs at
 emitted gluon energies $\sim 10 T$, such a treatment is necessary.  We
 find that the ratio $R$ of the data to the $pp$ based expectations at first
 falls with energy, reaches a minimum around $30 T$, and then rises slowly
 thereafter. 

 We have checked that $R$ approaches 1 at very large $p_T$.
 However, this happens at $p_T$ of a few hundred $T$.
 With a reasonable estimate of $T = 0.5\,\GeV$, this indicates that
 even at $p_T = 25\,\GeV$, the suppression pattern at LHC will be nearly flat.
 This is in contrast to the previous finding of 
 Gyulassy and Vitev~\cite{Vitev:2002pf}, who claim that
 the Cronin effect and nuclear shadowing are essential in describing 
 the flat $R$ from RHIC, but that these effects are small at the LHC, 
 which will therefore observe rising $R$ at large $p_T$.
 
 To turn our observations into a quantitative prediction for the $p_T$
 spectrum will require a proper inclusion of the nuclear geometry and of
 the expansion and cooling of the QGP medium.  We plan to address these
 issues in a future publication.


The authors thank C.~Gale and R.~Venugopalan for their comments and
suggestions.
S.J.~and G.M.~are supported in part by 
Natural Sciences and Engineering Research Council of Canada and by le Fonds
Nature et Technologies of Qu\'ebec.
S.J.~thanks 
RIKEN-BNL Center and the U.S. Department of Energy [DE-AC02-98CH10886] for
providing facilities essential for the completion of this work.

\end{document}